\documentclass[conference]{IEEEtran}
\IEEEoverridecommandlockouts
\usepackage{cite}
\usepackage{amsmath}
\usepackage[linesnumbered,ruled]{algorithm2e}
\usepackage{graphicx}
\usepackage{amssymb}
\usepackage{textcomp}
\usepackage{xcolor}
\def\BibTeX{{\rm B\kern-.05em{\sc i\kern-.025em b}\kern-.08em
    T\kern-.1667em\lower.7ex\hbox{E}\kern-.125emX}}

\begin{document}

\title{DNC-Aided SCL-Flip Decoding of Polar Codes\\
{\normalfont\large 
    Yaoyu Tao$^{1,2}$,~\IEEEmembership{Member,~IEEE,} and Zhengya Zhang$^{1}$,~\IEEEmembership{Senior~Member,~IEEE}%
  }\\[-1.3ex]}

\author{
    \IEEEauthorblockA{%
        $^1$Department of Electrical Engineering and Computer Science\\
        University of Michigan, Ann Arbor, MI 48109 USA\\
        Emails: \{taoyaoyu, zhengya\}@umich.edu\\[-5.0ex]
    }
    \and
    \IEEEauthorblockA{%
         $^2$Qualcomm Wireless R\&D \\San Jose, CA 95110 USA\\
         Emails: yaoyut@qti.qualcomm.com\\[-5.0ex]
    }
}

\maketitle

\begin{abstract}
Successive-cancellation list (SCL) decoding of polar codes has been adopted for 5G. However, the performance is not very satisfactory with moderate code length. Heuristic or deep-learning-aided (DL-aided) flip algorithms have been developed to tackle this problem. The key for successful flip decoding is to accurately identify error bit positions. In this work, we propose a new flip algorithm with help of differentiable neural computer (DNC). New state and action encoding are developed for better DNC training and inference efficiency. The proposed method consists of two phases: i) a flip DNC (F-DNC) is exploited to rank most likely flip positions for multi-bit flipping; ii) if decoding still fails, a flip-validate DNC (FV-DNC) is used to re-select error bit positions for successive flip decoding trials. Supervised training methods are designed accordingly for the two DNCs. Simulation results show that proposed DNC-aided SCL-Flip (DNC-SCLF) decoding demonstrates up to 0.34dB coding gain improvement or 54.2\% reduction in average number of decoding attempts compared to prior works.
\end{abstract}

\begin{IEEEkeywords}
Polar code, deep learning, successive cancellation list decoder, flip algorithms, differentiable neural computer
\end{IEEEkeywords}

\section{Introduction}
Capacity-achieving polar codes \cite{polar_arikan_first} have been adopted in modern communication systems such as 5th generation (5G) wireless standard. They can be decoded sequentially on a trellis using successive cancellation list (SCL) \cite{scl_tal_first} decoder. Upon receiving log-likelihood ratios (LLRs), SCL calculates path metrics (PMs) following a bit after bit order. A list of $L$ most likely paths are kept during decoding and decoded bits are determined by the most likely path that passes cyclic redundancy check (CRC). However, the decoding performance is not very satisfactory with moderate code length $N$. Once wrong bit decisions occur on the trellis, they have no chance to be corrected due to the sequential decoding order.

\begin{figure}
\centering
\includegraphics[width=.95\linewidth]{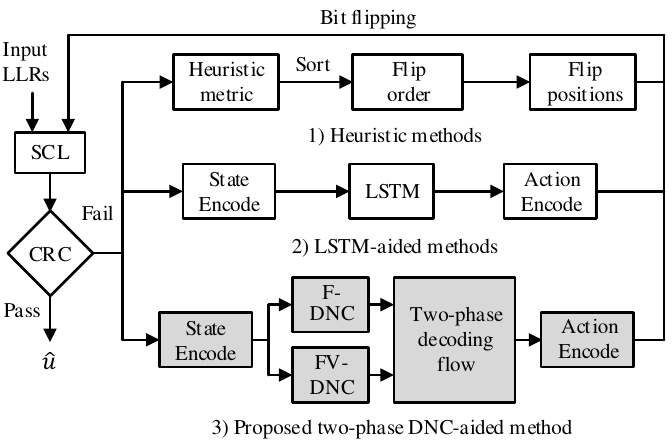}
\caption{Overview of 1) Heuristic bit flipping, 2) LSTM-aided bit flipping and 3) proposed DNC-aided two-phase bit flipping.}
\label{dnc_aided_sclf_top}
\end{figure}

To solve this problem, flip algorithms are used when standard decoding fails with CRC. Error bit positions are searched and flipped in subsequent decoding attempts. Clearly, the key for successful flip decoding is to accurately identify error bit positions. As shown in \figurename~\ref{dnc_aided_sclf_top}, heuristic methods \cite{polar_flip_first, bit_flip_gross_1, bit_flip_gross_2, polar_flip_critical_set, liu_lstm_scl, sclf_ieee_access, partitioned_flip, dynamic_flip_1, dynamic_flip_2, tao_polar_jssc, prac_dynamic_sclip_gross, fast_threshold_sclip_gross, error_dependency_scflip_huawei, generalized_sclf} use explicit mathematical metric to estimate the likelihood of each bit being an error bit. The likelihoods are sorted to obtain the flip position set. However, the optimal flipping strategy is still an open problem to date.

Recent works on flip algorithms involve deep learning (DL). DL-aided methods require state encoding to pre-process the inputs to the neural network (NN) and action encoding to generate flip position set from the NN outputs, as shown in \figurename~\ref{dnc_aided_sclf_top}. \cite{lstm_polar_flip_huawei, lstm_scl, lstm_polar_vtc2020, liu_lstm_scl} propose to use long short-term memory (LSTMs) to help locate flip positions for short polar codes of length 64 or 128. However, LSTMs lack the scalability to handle long-distance dependencies embedded in the sequential SCL decoding when code length increases, presenting a limitation for practical adoptions.

The recently developed differentiable neural computer (DNC) \cite{dnc_nature} addresses the scaling problem of LSTM. DNC can be considered as an LSTM augmented with an external memory through attention-based soft read/write mechanisms. In this paper, we propose to use DNC for bit flipping of practical-length polar codes to enhance the accuracy of identifying error bit positions. The main contributions are summarized as follows:

\begin{enumerate}
    \item A new two-phase decoding is proposed assisted by two DNCs, flip DNC (F-DNC) and flip-validate DNC (FV-DNC), as shown in \figurename~\ref{dnc_aided_sclf_top}. F-DNC ranks mostly likely flip positions for multi-bit flipping. If decoding still fails, FV-DNC is used to re-select flip positions for successive flip decoding trials. 
    \item We propose new action encoding with soft multi-hot scheme and state encoding considering both PMs and received LLRs for better DNC training and inference efficiency. Training methods are designed accordingly for the two DNCs, where training database are generated based on supervised flip decoding attempts.
    \item Simulation results show that the proposed DNC-aided SCL-Flip (DNC-SCLF) decoder outperforms the state-of-the-art techniques by up to 0.34dB in error correction performance or 54.2\% reduction in average number of decoding attempts.
\end{enumerate}

\section{Background}

\subsection{SCL Decoding of Polar Codes}

An ($N$, $K$) polar code has a code length $N$ and code rate $K/N$. Let $u^{N-1}_0 = (u_0, u_1, ..., u_{N-1})$ denote the vector of input bits to the encoder. The $K$ most reliable bits in $u^{N-1}_0$, called free bits, are used to carry information; while the remaining $N-K$ bits, called frozen bits, are set to pre-determined values. 

Successive cancellation (SC) \cite{polar_arikan_first} is the basic decoding scheme of polar codes. Assume $r^{N-1}_0$ is the received LLRs. It follows a bit-after-bit sequential order and calculates bit LLR $L^{\hat{u}_i}$ for $i$-th bit on the SC trellis, where $i = \{0,...,N-1\}$ and $\hat{u}_i = \pm 1$. The decoding of a bit depends on previously decoded bits. SC keeps the most likely path from the candidate paths at each bit level. SCL decoding \cite{scl_tal_first} improves the error-correction performance by keeping a list of $L$ mostly likely paths  through the PM values $\mathcal{P}(\ell)_i$, where $\ell$ and $i$ denote the path index and the bit index, respectively. For each path $\ell$ and each bit $i$, the PMs are defined as \eqref{pm_eqn}:

\begin{align}
\label{pm_eqn}
\begin{gathered}
\mathcal{P}(\ell)_i \triangleq \sum_{j = 0}^{i}{\text{ln}(1+e^{-(1-2\hat{u}_j(\ell))L^{\hat{u}_j}(\ell)})}
\end{gathered}
\end{align}

\noindent where $\hat{u}_j(\ell)$ and $L^{\hat{u}_j}(\ell)$ denote the $j$-th bit at $\ell$-th path and the bit LLR for $\hat{u}_j$ given received LLRs $r_0^{N-1}$ and decoding trajectories $\hat{u}_0^{j-1}(\ell)$, respectively. SC can be seen as a special case when list size $L = 1$. Concatenating polar code with CRC \cite{polar_crc_niu,polar_crc_li} can help pick the final path. 

\subsection{State-of-the-art Flip Algorithms}

Flip algorithms are proposed to identify error bit positions upon failed CRC. The flip positions can be determined by either heuristic metric or NNs like LSTMs. Heuristic methods like \cite{polar_flip_first,bit_flip_gross_1, bit_flip_gross_2, polar_flip_critical_set, partitioned_flip} use received LLRs or their absolute values as the metric to derive flip positions. Specifically, \cite{polar_flip_critical_set} introduces a critical set to reduce the search space of flip positions for lower complexity. \cite{partitioned_flip} subdivides the codeword into partitions, on which SC-Flip (SCF) is run for shorter latency. However, these methods can only flip one bit at a time. \cite{dynamic_flip_1,dynamic_flip_2,fast_threshold_sclip_gross,prac_dynamic_sclip_gross} propose a dynamic SC-Flip (DSCF) that allows flipping of multiple bits at a time and improves the latency of SCF. Multi-bit flipping requires identifying multiple error bit positions concurrently. DSCF introduces a new metric considering not only received LLRs but also the trajectories in the sequential SCL decoding. \cite{prac_dynamic_sclip_gross, fast_threshold_sclip_gross} introduce variations of DSCF to improve the accuracy of identifying error bit positions. \cite{sclf_ieee_access, generalized_sclf} extends the bit-flipping from SC to SCL for a SCL-Flip decoding (SCLF). Similarly, SCF is a special case of SCLF when $L=1$. 

Recently developed DL-aided SCF/SCLF \cite{lstm_polar_flip_huawei,lstm_polar_vtc2020,liu_lstm_scl,lstm_scl} exploit a trained LSTM to locate error bit positions instead of heuristic metric. They have shown slightly better performance than heuristic methods for short polar codes of length 64 or 128. However, the accuracy of identifying error bit positions is limited by the scalability of LSTMs when code length increases. On the other hand, state-of-the-art LSTM methods use simple state and action encoding that do not support multi-bit flipping efficiently, resulting in more decoding attempts compared to heuristic methods. 

\subsection{Differentiable Neural Computer (DNC)}

\begin{figure}
\centering
\includegraphics[width=.83\linewidth]{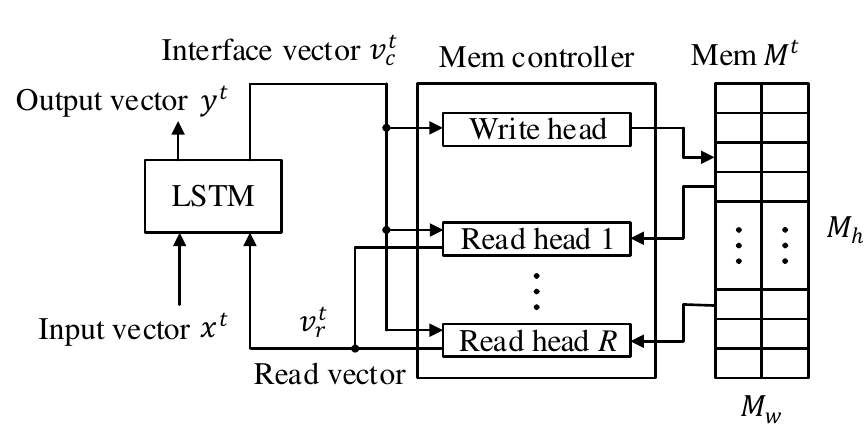}
\caption{Top-level architecture of DNC.}
\label{dnc_arch}
\end{figure}

DNC addresses LSTM's scalability problem with help of an external memory. Since its invention, DNC has found many applications like question answering \cite{NIPS2015_5846, pmlr_kumar16}. DNC can be considered as an LSTM augmented with an external memory through soft read and write heads, as shown in \figurename~\ref{dnc_arch}. In this work, we use DNCs to enhance the accuracy of identifying error bit positions.

A top level architecture of DNC is demonstrated in \figurename~\ref{dnc_arch}. DNC periodically receives $x^t$ as input vector and produces $y^t$ as output vector at time $t$. The output vector $y^t$ is usually made into a probability distribution using softmax. At time $t$, the DNC 1) reads an input $x^t$, 2) writes the new information into the external memory using interface vector $v_c^t$ through memory controller, 3) reads the updated memory $M^t$ and 4) produces an output $y^t$. Assume the external memory is a matrix of $M_h$ slots, each slot is a length-$M_w$ vector. To interface with this external memory, DNC computes read and write keys to locate slots. The memory slot is found using similarity between key and slot content. This mechanism is known as the content-based addressing. In addition, DNC also uses dynamic memory allocation and temporal memory linkage mechanisms for computing write and read weights. We omit the mathematical descriptions of DNC here and readers can refer to \cite{dnc_nature} for more details. 

\section{DNC-Aided Flip Decoding}
\label{dnc_for_sclf_section}

Bit-flipping can be modeled as a game and the DNC is the player to identify flip positions towards successful decoding. Upon CRC failure, the DNC player needs to take an action based on current state, either reverting falsely flipped positions or adding more flip positions. The proposed DNC-aided method includes: 1) new state and action encoding; and 2) a DNC-aided two-phase decoding flow. 

\subsection{State and Action Encoding}

One of the keys for efficient DNC inference is to design good input (state) and output (action) vector for training and inference. We discuss the encoding of existing LSTM-based approaches \cite{lstm_polar_flip_huawei, liu_lstm_scl, lstm_polar_vtc2020, lstm_scl} and present a new encoding scheme.

\subsubsection{State Encoding}
a straightforward way to encode states is to directly use the received LLR sequence $r^{N-1}_0$. \cite{lstm_polar_flip_huawei,liu_lstm_scl} use the amplitudes of received LLRs as the LSTM input. \cite{lstm_polar_vtc2020} uses the amplitudes of received LLRs combining the syndromes generated by CRC for state encoding. However, path metric information in sequential decoding are discarded in these methods, resulting in a loss in representing error path selection probability. \cite{lstm_scl} proposed a state encoding by taking the PM ratio of discarded paths and survival paths. However, this representation requires extra computations for PM summations at each bit position and does not include received LLR information.

In this work, we introduce a new state encoding scheme using the gradients of $L$ survival paths concatenated with received LLRs. It takes both PMs and received LLRs into consideration. The PM gradients $\triangledown \mathcal{P}(\ell)_i$ for $i$-th bit can be described as \eqref{pm_gradient}:

\begin{align}
\label{pm_gradient}
\triangledown \mathcal{P}(\ell)_i = \text{ln}(1+e^{-(1-2\hat{u}_i(\ell))L^{\hat{u}_i}(\ell)})
\end{align}

Note that $\triangledown \mathcal{P}(\ell)_i$ can be directly taken from existed PM calculations in standard SCL without extra computations. The state encoding $S$ is therefore a vector as \eqref{dnc_state_encoding} and is used as DNC input in this work.

\begin{align}
\label{dnc_state_encoding}
S = \{\triangledown \mathcal{P}(\ell)_0^{N-1}, r_0^{N-1}\}
\end{align}

\subsubsection{Action Encoding} the one-hot scheme used in state-of-the-art LSTM-based flip algorithms are efficient in identifying the first error bit, but lacks the capability to flip multiple bits at a time. This results in more decoding attempts. To improve bit flipping efficiency, we propose a soft multi-hot (i.e. $\omega$-hot) flip vector $v_f$ to encode both first error bit and subsequent error bits, aiming to correctly flip multiple bits in one attempt. $v_f$ is a length-$N$ vector that has $\omega$ non-zero entries. An action is therefore encoded by $v_f$. Each possible flip position in $v_f$ is a non-zero soft value indicating the flip likelihood of the bit. 

For training purpose, we introduce a scaled logarithmic series distribution (LSD) to assign flip likelihoods to the $\omega$ flip positions, where $p\in(0,1)$ is a shape parameter of LSD. The intention is to create a distribution with descending probabilities for first error bit position and subsequent error bit positions and to provide enough likelihood differences between them. Suppose the $k$-th bit in polar code has an index $\mathcal{I}_\mathcal{F}(k)$ in the flip position set $\mathcal{F}$. Non-zero entries of $v_f$ can be derived as \eqref{vf_ref}: 

\begin{align}
\label{vf_ref}
\begin{gathered}
v_f(k) = \mathcal{K}\frac{-1}{\text{ln}(1-p)}\frac{p^{\mathcal{I}_\mathcal{F}(k)}}{\mathcal{I}_\mathcal{F}(k)} \text{ for } k \in \mathcal{F} \\
\text{where scaling factor } \mathcal{K} = 1/\int_\mathcal{F}v_f
\end{gathered}
\end{align}

Reference $v_f$ generation for training are discussed in Section~\ref{fdnc_train}. The impacts of parameters $\omega$ and $p$ on the accuracy of identifying error bit positions are discussed in Section~\ref{param_analysis}.. 

\subsection{DNC-Aided Two-Phase Decoding Flow} 
\label{dnc_flip_decode_flow}

\begin{figure}
\centering
\includegraphics[width=.95\linewidth]{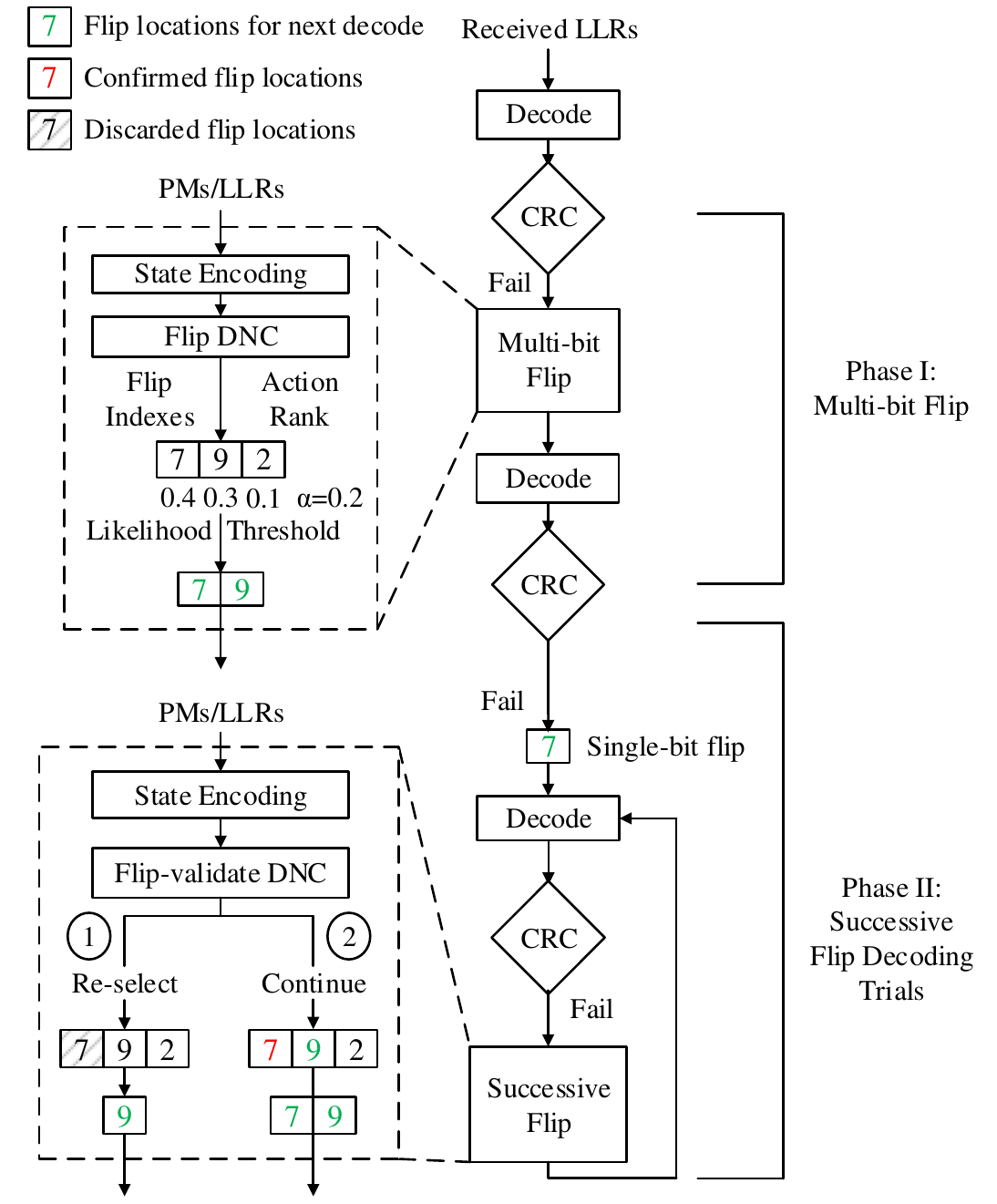}
\caption{DNC-aided two-phase flip decoding ($\omega = 3$ case).}
\label{dnc_flip_flow}
\end{figure}

We design a two-phase flip decoding flow aiming to reduce the number of SCL attempts while achieving good error correction performance. The two phases in this flow are: i) multi-bit flipping and ii) successive flip decoding trials. In the first phase, the received symbols are first decoded with a standard decoder. If it fails CRC, a flip DNC (F-DNC) exploits the state encoding $S$ to score the actions, i.e., estimate the probability of each bit being error bits and output a flip vector $v_f$. \figurename~\ref{dnc_flip_flow} shows an example of $\omega = 3$ where $\mathcal{F} = \{7, 9, 2\}$ is the flip position set with descending likelihoods \{0.4, 0.3, 0.1\}. To avoid wrong flips of subsequent positions with insignificant flip likelihoods, an $\alpha$-thresholding is applied to keep only positions with $v_f(i)>\alpha, i =\{ 0,...,N-1\}$, for multi-bit flipping. A subsequent decode attempt is then carried out with multi-bit flipping of bit positions $\{7,9\}$ in the example.

If CRC still fails after multi-bit flipping, we enter Phase-II that successively re-select or confirm a single error bit position. The reasons of failed decoding in Phase-I are either: 1) first error bit position is wrong; or 2) first error bit position is right but some subsequent flip positions are wrong. Our proposed solution is to flip each possible error bit position one at a time and use a flip-validate DNC (FV-DNC) to confirm if this is a correct flip before moving to the next possible error bit position. The first attempt in Phase-II flips the highest ranked error bit position in $\mathcal{F}$, i.e., bit $7$ in the example shown in \figurename~\ref{dnc_flip_flow}.

\begin{figure}
\centering
\includegraphics[width=.7\linewidth]{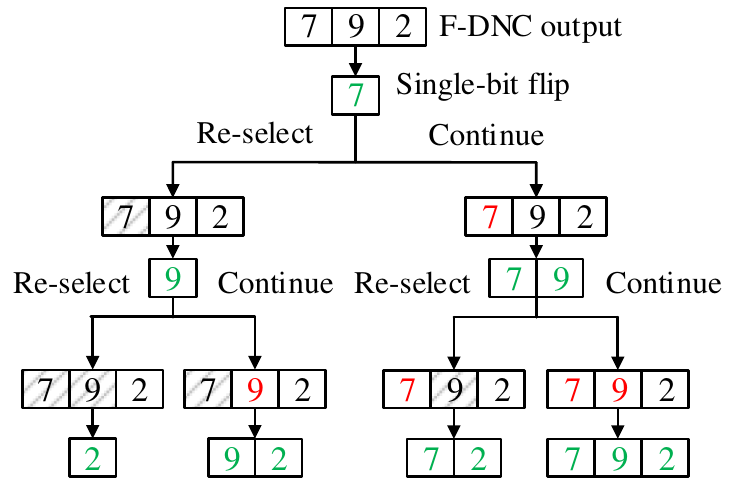}
\caption{Flip attempts in Phase-II for different FV-DNC output combinations ($\omega = 3$ case).}
\label{phase_2_tree}
\end{figure}

If FV-DNC invalidates the single-bit flip (bit $7$ in this case), we discard bit $7$ and re-select the flip position to next bit $9$ in $\mathcal{F}$. Alternatively, if FV-DNC confirms the flip of bit $7$, we continue by adding bit $9$ into the flip queue $\mathcal{Q}_f$ and flip $\mathcal{Q}_f = \{7, 9\}$ in next attempt. The process runs successively until CRC passes or reaching the end of $\mathcal{F}$. \figurename~\ref{phase_2_tree} shows all possible flip combinations given different FV-DNC output combinations in the $\omega = 3$ case. The number of decoding attempts of Phase-II is bounded by $\omega$. The two-phase DNC-SCLF can be described as Algorithm~\ref{alg:dnc_polar_flip}. 

\begin{algorithm}
$\hat{u}_0^{N-1},S \leftarrow\text{ SCL}(r_0^{N-1})$ \\
\textbf{if} $\text{CRC}(\hat{u}_0^{N-1}) = \text{pass}$\textbf{ return} $\hat{u}_0^{N-1}$ 

\textbf{Phase-I}: Multi-bit Flipping \\
$\mathcal{F}, \omega, v_f \leftarrow \text{F-DNC}(S)$ \\
$\hat{u}_0^{N-1}\leftarrow\text{ SCL}(r_0^{N-1}, \mathcal{F}_{v_f\geq \alpha})$ \\
\textbf{if} $\text{CRC}(\hat{u}_0^{N-1}) = \text{pass}$\textbf{ return} $\hat{u}_0^{N-1}$\

\textbf{Phase-II}: Successive Flip Decoding Trials \\
$\mathcal{Q}_f = \{\mathcal{F}[0]\}$ \\
\For{$i = 0,1,...,\omega-1$}{
    $\hat{u}_0^{N-1},S \leftarrow\text{ SCL}(r_0^{N-1}, \mathcal{Q}_f)$ 
    
    $\textbf{if } \text{CRC}(\hat{u}_0^{N-1}) = \text{pass or }i = \omega-1 \textbf{ return } \hat{u}_0^{N-1}$ \\
    $\mathcal{R} \leftarrow \text{FV-DNC}(S)$ \\
  \eIf{$\mathcal{R} = \text{\normalfont continue}$}{
    $\mathcal{Q}_f = \{\mathcal{Q}_f, \mathcal{F}[i+1]\}$
   }{
    $\mathcal{Q}_f[\text{end}] = \mathcal{F}[i+1]$
  }
 }
\caption{DNC-Aided SCL-Flip Decoding}
\label{alg:dnc_polar_flip}
\end{algorithm}

\section{Training Methodology}
\label{fdnc_train}

In this section, we discuss training for the DNCs used in proposed DNC-SCLF. The training is conducted off-line and does not increase the run-time decoding complexity. We adopt the cross-entropy function which has been widely used in classification tasks \cite{lecun_nature_dl}. 

\subsection{F-DNC Training}

In the first training stage, we run extensive SCL decoder simulations and collect error frames upon CRC failure. The F-DNC training database consists of pairs of $S$ from \eqref{dnc_state_encoding} as DNC input and a corresponding $v_f$ from \eqref{vf_ref} as reference output. $S$ can be straightforwardly derived based on received LLRs and PMs of collected error frames. However, $v_f$ is determined by parameter $\omega$ and $p$, whose values will affect the training and inference efficiency. We first label the error bit positions w.r.t the transmitted sequence for each sample as candidate flip positions. Intuitively, small $\omega$ and $p$ strengthen the likelihood of identifying first error bit position, but attenuate the likelihoods of subsequent error bit positions. Hence there is a trade-off between the accuracy of identifying first error bit position and the accuracy of identifying subsequent error bit positions. In this work, we carried out reference $v_f$ generations with $\omega = \{2, 5, 10\}$ and $p = \{0.2, 0.8\}$. The experimental results with these parameter choices are discussed in Section~\ref{experiment_analysis}.

\begin{table}
	\centering
	\caption{F-DNC/FV-DNC Hyper-parameters Set}
	\renewcommand{\arraystretch}{1.2}
	\begin{tabular}{|c|c|}
        \hline
		Parameter & Description \\ \hline
		LSTM controller & 1 layer of size 128 \\ \hline
		Size of access heads & 1 write head, 4 read heads\\ \hline
		Size of external memory & $M_h = 256, M_w = 128$ \\ \hline
		Size of training set & $10^6$ for F-DNC, $3\times10^7$ for FV-DNC \\ \hline
	    Size of validation set & $5\times10^4$ \\ \hline
		Mini-batch size & 100 \\ \hline
		Dropout probability & 0.05 \\ \hline
		Optimizer & Adam \\ \hline
		Environment & Tensorflow 1.14.0 on Nvidia GTX 1080Ti \\ \hline 
	\end{tabular}
	\label{tbl:fdnc_params}
\end{table}

\subsection{FV-DNC Training}

The error frames that can not be decoded correctly in Phase-I enter Phase-II, where single bit positions are flipped and tested successively as shown in \figurename~\ref{phase_2_tree}. This is to prevent wrong flips that will lead the DNC player into a trapping state and can never recover. The FV-DNC is a classifier taking either "re-select" or "continue" action given the knowledge of received LLRs and PMs from most recent attempt. The key for FV-DNC training is to create a well-categorized database that can detect trapping state effectively. We carry out supervised flip decoding attempts based on reference $v_f$ in F-DNC database. For each collected error:1) the first 5 error bit positions in reference $v_f$ are flipped bit after bit and their corresponding state encoding $S$ are recorded. These samples result in a ``continue'' action. 2) After flipping each of the first 5 error bit positions, we flip 5 random positions and record their state encoding $S$. These samples indicate trapping state and result in a ``re-select'' action. For each collected frame, we have 5 samples for ``continue'' action and 25 samples for ``re-select'' action. 

\section{Experiments and Analysis}
\label{experiment_analysis}

To show the competitiveness of DNC in tackling long-distance dependencies in polar decoding trellis, we evaluate the performances for polar codes of length $N$ = 256, 1024 with SC and SCL ($L=4$). The code rate is set to 1/2 with an 16b CRC. Error frames are collected at SNR 2dB. In this paper we do not focus on the hyper-parameter optimizations for DNC and just demonstrate a set of configurations that work through our experiments for F-DNC and FV-DNC in Table~\ref{tbl:fdnc_params}. 

\begin{figure}
\centering
\includegraphics[width=0.98\linewidth]{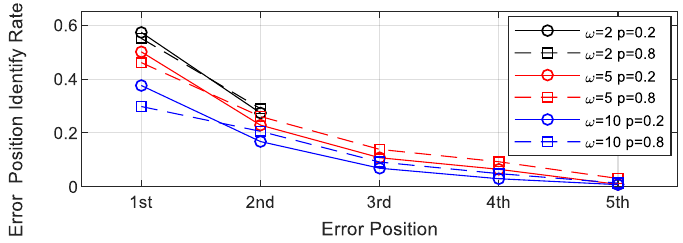}
\caption{Rate of identifying error bit positions for $\omega=\{2,5,10\}$ and $p=\{0.2,0.8\}$ for SC decoding of (256,128) polar code.}
\label{param_study}
\end{figure}

\begin{figure}
\centering
\includegraphics[width=0.98\linewidth]{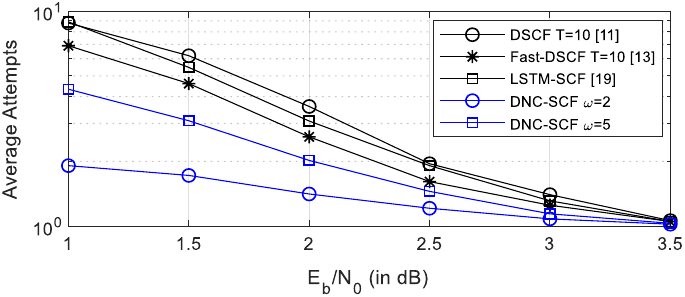}
\caption{Number of extra decoding attempts of DNC-SCF and state-of-the-art flipping algorithms for (1024, 512) polar code.}
\label{dnc_sclf_latency}
\end{figure}

\subsection{Accuracy of Identifying Error Bits}
\label{param_analysis}

Firstly, we study the impacts of parameters $\omega$ and $p$ introduced in action encoding. For a fair comparison, we pick the same code length $N=256$ and SC decoding used in heuristic method \cite{dynamic_flip_2} and LSTM-based method \cite{lstm_polar_flip_huawei}. \figurename~\ref{param_study} presents the accuracy of identifying the first 5 error bit positions. For a given $\omega$, a smaller $p$ ($p=0.2$) enhances the probability of identifying the first error bit position, but attenuates the probability of identifying subsequent error bit positions. We achieve a 0.573 success rate of identifying the first error bit position with $\omega=2$, outperforming the 0.425 and 0.51 success rate with heuristic DSCF \cite{dynamic_flip_2} and LSTM-aided SCF \cite{lstm_polar_flip_huawei}, respectively. Comparing $\omega = 2$ and $\omega = 5$ with same $p=0.8$, a bigger $\omega$ helps to identify more error bit positions, but the success rates of identifying each position are degraded.

We pick $p=0.8$ in our two-phase DNC-SCLF experiments to strengthen the success rates of identifying subsequent error bit positions and slightly sacrifice the success rate of identifying first error bit position. This is because with help of FV-DNC, even though F-DNC may not identify the first error bit position accurately in Phase-I, the two-phase decoding can re-select it in Phase-II. We use an $\alpha=0.03$ for thresholding through our experiments. 

\subsection{Complexity and Latency}

Metric calculation and sorting in heuristic methods can be implemented inside standard SC/SCL decoders. However, DL-aided algorithms introduce higher complexity and require an inference accelerator to interact with the decoder. We use GPU that achieves a speed of 1.7 ms/inference. For practical adoptions, a dedicated accelerator can be implemented for faster inference. Bit flipping is conditionally triggered when the standard decoder fails and the triggering rate is lower than the FER. DL-aided algorithms are more suitable for the low FER regime where the inference latency can be hidden behind successful decoding runs with help of LLR buffers. In this work we do not focus on the inference acceleration and LLR buffering strategy, but focus on the average number of flip decoding attempts that determines the overall latency. 

Assume $\beta_{1}$ is the rate of successful decoding with multi-bit flipping in Phase-I, the average number of decoding attempts $T_{avg}$ for a DNC-aided flip decoding can be calculated as \eqref{tavg_eqn}:

\begin{align}
\label{tavg_eqn}
T_{avg} = \beta_{\text{1}} + \omega_{\text{2,avg}}(1-\beta_{\text{1}})
\end{align}

\noindent where $\omega_\text{2,avg}$ is the average number of attempts in Phase-II and $\omega_\text{2,avg} \leq \omega$. \figurename~\ref{dnc_sclf_latency} demonstrates the $T_{avg}$ for proposed DNC-SCF and the state-of-the-art techniques. At a 2dB SNR, DNC-SCF with $\omega = 2$ improves the average decoding attempts by 45.7\% and 54.2\% compared to state-of-the-art heuristic \cite{prac_dynamic_sclip_gross} and LSTM-aided methods \cite{lstm_polar_vtc2020}, respectively.

\subsection{Error-Correction Performance}

We compare coding gain of DNC-SCF at FER 10$^{-4}$ with state-of-the-art heuristic methods \cite{dynamic_flip_2,prac_dynamic_sclip_gross} and LSTM-based methods \cite{lstm_polar_vtc2020} for a (1024, 512) polar code and 16b CRC. DNC-SCF $\omega=2$ achieves 0.5dB coding gain w.r.t SC decoder. Increasing $\omega$ to 5 provides another 0.31dB coding gain. DNC-SCF $\omega=5$ also outperforms DSCF \cite{dynamic_flip_2} or Fast-DSCF \cite{prac_dynamic_sclip_gross} with $T=10$ by 0.03dB and 0.05dB, respectively, while reducing the number of decoding attempts by 45.7\%. Further increasing $\omega$ to DNC-SCF $\omega=10$ provides 0.21dB coding gain compared to DSCF $T=10$ while reducing the number of decoding attempts by 18.9\%. 

\begin{figure}
\centering
\includegraphics[width=\linewidth]{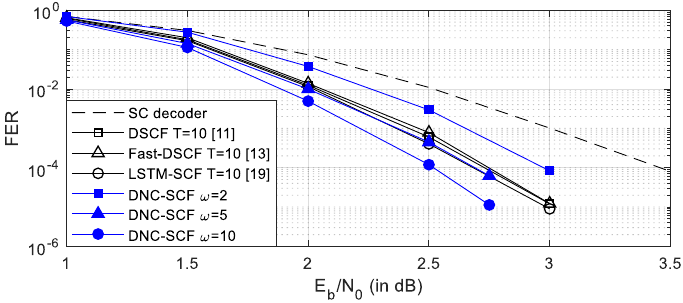}
\caption{FER performance comparison of DNC-SCF and state-of-the-art flipping algorithms for (1024,512) polar code and 16b CRC.}
\label{dnc_scf_dscf_fer}
\end{figure}

\begin{figure}
\centering
\includegraphics[width=\linewidth]{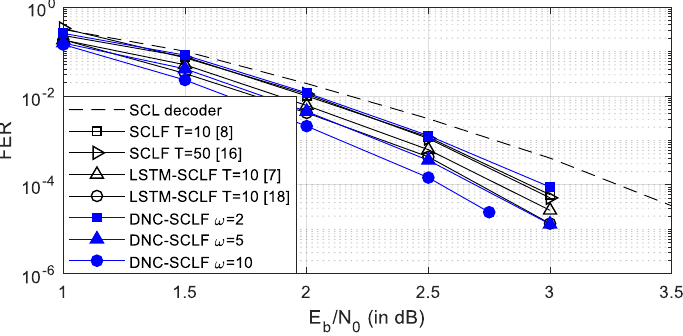}
\caption{FER performance comparison of DNC-SCLF ($L=4$) and state-of-the-art flipping algorithms for (256,128) polar code and 16b CRC.}
\label{dnc_sclf_fer_cpm}
\end{figure}

The LSTM-based approach in \cite{lstm_polar_flip_huawei} does not report FER, but has shown up to 10\% improvement in the accuracy of identifying first error bit position over DSCF with $T=1$ at 1dB SNR for (64, 32) polar code. Another LSTM-based SCF \cite{lstm_polar_vtc2020} provides FER for (64, 32) polar code with $T = 6$ and claims 0.2dB improvement over DSCF $T=6$. The FER of \cite{lstm_polar_vtc2020} with 1024b and $T=10$ is shown in \figurename~\ref{dnc_scf_dscf_fer}, worse than DNC-SCF $\omega=5$. LSTM's capability of identifying error bit positions gets weakened when code length increases.

We further compare the FER of DNC-SCLF ($L=4$) on (256, 128) polar code and 16b CRC with state-of-the-art heuristic methods \cite{sclf_ieee_access, generalized_sclf} and LSTM-based approaches \cite{liu_lstm_scl, lstm_scl} as shown in \figurename~\ref{dnc_sclf_fer_cpm}. At FER $10^{-4}$, DNC-SCLF $\omega=2$ achieves a 0.27dB coding gain w.r.t standard SCL. Increasing $\omega$ to 5 results in 0.59dB coding gain from the standard SCL. DNC-SCLF $\omega=5$ achieves 0.21dB and 0.01dB better performance than heuristic SCLF \cite{generalized_sclf} and LSTM-SCLF \cite{lstm_scl} with $T=10$, respectively. Further increasing $\omega$ to DNC-SCLF $\omega=10$ improves the coding gain to 0.34dB and 0.16dB compared with \cite{generalized_sclf} and \cite{lstm_scl}, respectively.

\section{Conclusions}
In this paper, we present a new DNC-aided SCLF decoding. We propose a two-phase decoding assisted by two DNCs, F-DNC and FV-DNC, to identify error bit positions for multi-bit flipping and to re-select error bit positions for successive flip decoding trials, respectively. The multi-bit flipping reduces number of flip decoding attempts while successive flip decoding trials lowers the probability of going into trapping state. Training methods are proposed accordingly to efficiently train F-DNC and FV-DNC. Simulation results show that the proposed DNC-SCLF helps to identify error bits more accurately, achieving better error correction performance and reducing the number of flip decoding attempts than the the state-of-the-art flip algorithms. We plan to investigate the parameter optimizations for proposed DNC-SCLF in follow-up research.

\bibliographystyle{IEEEtran}
\bibliography{IEEEabrv,icc2021}

\end{document}